\documentclass[summary]{gass2017}
\usepackage{amsmath,amsfonts,graphicx,array,multirow,color}
\usepackage{graphicx}
\usepackage[export]{adjustbox}
\usepackage[caption=false]{subfig}
\usepackage{enumitem}

\title{Comparison of signal detectors for time domain radio SETI}
\author{Gregory Hellbourg\affref{ref1}, Andrew Xu\affref{ref2}}

\affiliation{
\aff{ref1}{Berkeley SETI Research Center - University of California - Berkeley, CA, USA}
\aff{ref2}{Department of Electrical Engineering and Computer Sciences - University of California - Berkeley, CA, USA}
}

\begin{document}

\maketitle

\begin{abstract}
The radio Search for Extra Terrestrial Intelligence (SETI) \cite{ref:seti2020} aims at identifying intelligent and communicative civilizations in the Universe through the detection of
engineered transmissions.
%signals of unknown nature and origin (signaling beacon or radio leakage).
In the absence of prior knowledge concerning the expected signal, SETI detection pipelines necessitate high sensitivity, versatility, and limited computational complexity to maximize the search parameter space and minimize the probability of misses.
This paper addresses the SETI detection problem as a binary hypothesis testing problem, and compares four detection schemes exploiting artificial features of the data collected by a single receiver radio telescope. After a theoretical comparison, those detectors are applied to real data collected with the Green Bank Telescope in West Virginia (USA).
\end{abstract}

\section{Introduction}

The radio Search for Extra Terrestrial Intelligence (SETI) aims at detecting radio transmissions
from intelligent and communicative civilizations in the Universe. High sensitivity radio telescopes receivers and detection schemes, associated to dedicated voltage data recorders, are necessary to maximize the searched parameter space (directions-of-arrival, epochs, frequency, signal characteristics...) and the probability of detection. The absence of prior information concerning the signal-of-interest also necessitates the exploration various signal features to minimizes the probability of misses.

This paper addresses artificial signal detection for single receiver radio telescopes. Section \ref{sec:signal_detection} introduces the SETI detection problem from a binary hypothesis testing standpoint. Four detectors are compared onto three data models in this study, all are described in this same section. Section \ref{sec:simulation} details the Monte-Carlo simulation ran to achieve the comparison, and draws the main conclusions. Section \ref{sec:realdata} analyzes the results of detectors applied to real telescope data. Section \ref{sec:conclusion} finally concludes this study and suggests future research directions to improve the reliability and computational cost of a single receiver radio SETI experiment.

\section{SETI signal detection}
\label{sec:signal_detection}

\subsection{Detection theory}

The SETI detection problem is formulated as a binary hypothesis testing problem \cite{ref:detection}, with the following two hypotheses $\emph{H0}$ and $\emph{H1}$:

\begin{equation}
\mathbf{x}[n] = \begin{cases}
\mathbf{x}_{\text{noise}}[n] & (\emph{H0})\\
\mathbf{x}_{\text{ET}}[n] + \mathbf{x}_{\text{noise}}[n] & (\emph{H1})
\end{cases}
\end{equation}
where $\mathbf{x}[n]$ is the basebanded, digitized, and channelized telescope output, $\mathbf{x}_{\text{noise}}[n]$ is the system noise contribution and $\mathbf{x}_{\text{ET}}[n]$ the eventual extra terrestrial (ET) transmission.

Given a detector $\theta(\mathbf{x}[n])$ and threshold $\gamma$, the detection performance is assessed through the evaluation of its Probability of Detection $P_d(\gamma)$ and Probability of False Alarm $P_{fa}(\gamma)$ \cite{ref:detection}.%, defined as:

%\begin{equation}
%P_d = \int_{\gamma}^{\infty} \mathcal{P}[\theta(\mathbf{x}[n])|H1] d\theta \ \ \ \ 
%P_{fa} = \int_{\gamma}^{\infty} \mathcal{P}[\theta(\mathbf{x}[n])|H0] d\theta
%\end{equation}

\subsection{Proposed detectors}
\label{sec:detectors}

Assume $N$ samples are received. The following detectors are proposed for comparison for a single dish telescope radio SETI experiment.

\begin{itemize}
\item Energy detection (Runtime: $\Theta (N)$):\\
The energy detector is defined as:\\
\begin{equation}
\mathbf{\theta}_{\text{energy}} = \frac{1}{N} \sum_{n=1}^N \mathbf{x}[n]\mathbf{x}^{*}[n]
\end{equation}\\
with $(.)^{*}$ the complex conjugate operator. The energy being conservative, this detector monitors the increase of energy contributed by an additional signal component.

\item Fourier-based detection (Runtime: $\Theta (\tilde{N} \text{log} \tilde{N})$):\\
The averaged periodogram detector the time series $\mathbf{x}[n]$ is defined as:\\
\begin{equation*}
\mathbf{\theta}_{\text{FT}} = \max_{f \in [1..\tilde{N}]} \frac{1}{M}\sum_{m=1}^{M}\left|\mathcal{F}_{\mathbf{x}[(m-1).\tilde{N}+1:m.\tilde{N}]}(f)\right|^2
\end{equation*}\\
where $\mathcal{F}_{\mathbf{x}[m.\tilde{N}:(m+1).\tilde{N}]}(f)$ is the Fourier transform of the $m^{\text{th}}$ non-overlapping block of $\tilde{N}$ samples of $\mathbf{x}[n]$ such that $N = M\times\tilde{N}$, and $\left|.\right|$ is the absolute value operator. This detector monitors spectral features embedded in the received time series.

%\subsubsection{Hamming-Fourier-based detection}
%
%This detector is an alternative to the \emph{Fourier-based detection}, where the non-overlapping blocks of $\tilde{N}$ samples of $\mathbf{x}[n]$ are multiplied by a Hamming window prior to applying the Fourier transform. The Hamming window reduces the spectral leakage of the simple Fourier transform. This window is often used for narrowband component detection \cite{ref:hamming}.

%Runtime: $\Theta (\tilde{N} \text{log} \tilde{N})$
\item Time-lag-based detection (Runtime: $\Theta (N \text{log} N)$):\\
The time-lag detector is expressed as:\\
\begin{equation}
\mathbf{\theta}_{\tau} = \max_{0 < \tau < N} \frac{R_{xx^*}(n,\tau)}{\mathbf{\theta}_{\text{energy}}}
\end{equation}\\
where $R_{xx^*}(n,\tau) = \frac{1}{N-\tau-1} \sum_{n=\tau+1}^{N} \mathbf{x}[n]\mathbf{x}^{*}[n-\tau]$ is the autocorrelation of $\mathbf{x}[n]$ at discrete time step $\tau\in\mathbb{Z}$. This detector monitors the whiteness of the received time series.

\item Karhunen-Lo\`eve Transform (Runtime: $\Theta {M^3}$):\\
The Karhunen-Lo\`eve Transform detector \cite{ref:klt} extends the \emph{Time-lag-based detection}, and is based on the eigen value decomposition of the auto-correlation matrix $\mathbf{R}$ of the received time series $\mathbf{x}[n]$:\\
\begin{equation}
\mathbf{R} = \frac{1}{N-M-1}\sum_{n=1}^{N-M}\mathbf{x}[n:n+M]\mathbf{x}^H[n:n+M]  % \mathcal{E}\{\mathbf{x}[n]\mathbf{x}^H[n]\}
\end{equation}
with $(.)^H$ the Hermitian transpose operator. The \emph{Karhunen-Lo\`eve Transform detector (KLT)} is expressed as:\\
\begin{equation}
\mathbf{\theta}_{KLT} = \max_{0 < m < M} \lambda_m / \text{trace}(\mathbf{R})
\end{equation}
with $\lambda_m$ the $m^{\text{th}}$ eigenvalue of $\mathbf{R}$, and $\text{trace}(.)$ the matrix trace operator.

The rank of $\mathbf{R}$ quantifies the trade-off between its integration time and its dimensionality, and is the main parameter of the detector. Large dimensionality is necessary for detecting high dimensional signal subspaces \cite{ref:mdl}, while long integration time ensures the estimation quality.
\end{itemize}

\subsection{Data models}
\label{sec:datamodels}
Three data models are considered to compare the performances of the proposed detectors.

\begin{itemize}
\item Noise model\\
The system noise includes independent contributions of the cosmic background, stellar responses, and from the instrumentation itself. It is modeled as a white circular stationary (over short durations) centered complex, independently and identically distributed, Gaussian process with power : $\mathbf{x}_{\text{noise}}[n] \sim \mathcal{NC}(0,\sigma^2)$

\item Linear chirp\\
A narrow-band ET transmission undergoes a frequency drift resulting from the Doppler effect due to the Earth rotation.
%q, continuously affecting the relative distance between emitter and receiver.
Frequency drifts are assumed linear over short observation durations ($\sim$ 5 minutes), and might reach up to a few Hz/s \cite{ref:targetselection}. The resulting signal model follows $\mathbf{x}_{\text{ET}}[n] = A.\text{exp}\{i 2 \pi \left[d.n + f_0\right] n\}\text{exp}\{i 2 \pi \phi\}$,
with power $\sigma_{\text{chirp}}^2 = A^2$, and $d/2\in\mathbb{R}$ the linear chirp rate.

\item Oversampled Binary Phase-Shift Keying (BPSK)\\
Information-bearing transmissions are likely, although the trade-off between detectability and channel capacity remains argued in SETI. A Binary Phase-Shift Keying (BPSK) modulation scheme has been selected for this study, and is modeled following: $\mathbf{x}_{\text{ET}}[n] = \xi_m . A .\text{exp}\{i 2 \pi f_0 n\}\text{exp}\{i 2 \pi \phi\}$, with power $\sigma_{\text{BPSK}}^2 = A^2$, and $\xi_m$ a pseudo-random variable taking value in \{-1,+1\}. The baud-rate is assumed lower than the sampling frequency. The discrete message $\xi_m$ can be convolved with a Hamming window for spectral smearing control (referred to as windowed BPSK).
\end{itemize}

\section{Detector comparison on simulated data}
\label{sec:simulation}

\subsection{Simulation setup}

The detectors proposed in section \ref{sec:detectors} are compared on the data models described in section \ref{sec:datamodels} through a Monte Carlo simulation involving $10^3$ independent trials per Signal-to-Noise Ratio (SNR) defined as $\text{SNR} = A^2/\sigma^2$. Two data sets of length $N = 2^{16}$ are generated for each trial, corresponding to the two hypothesis \emph{H0} and \emph{H1}. The statistics of each detector are then evaluated under both hypotheses.

\begin{figure}[t]
\centering
%\subfloat[]{\includegraphics[clip,width=.8\columnwidth]{./figures/sin}}\\
\subfloat[]{\includegraphics[clip,width=.8\columnwidth]{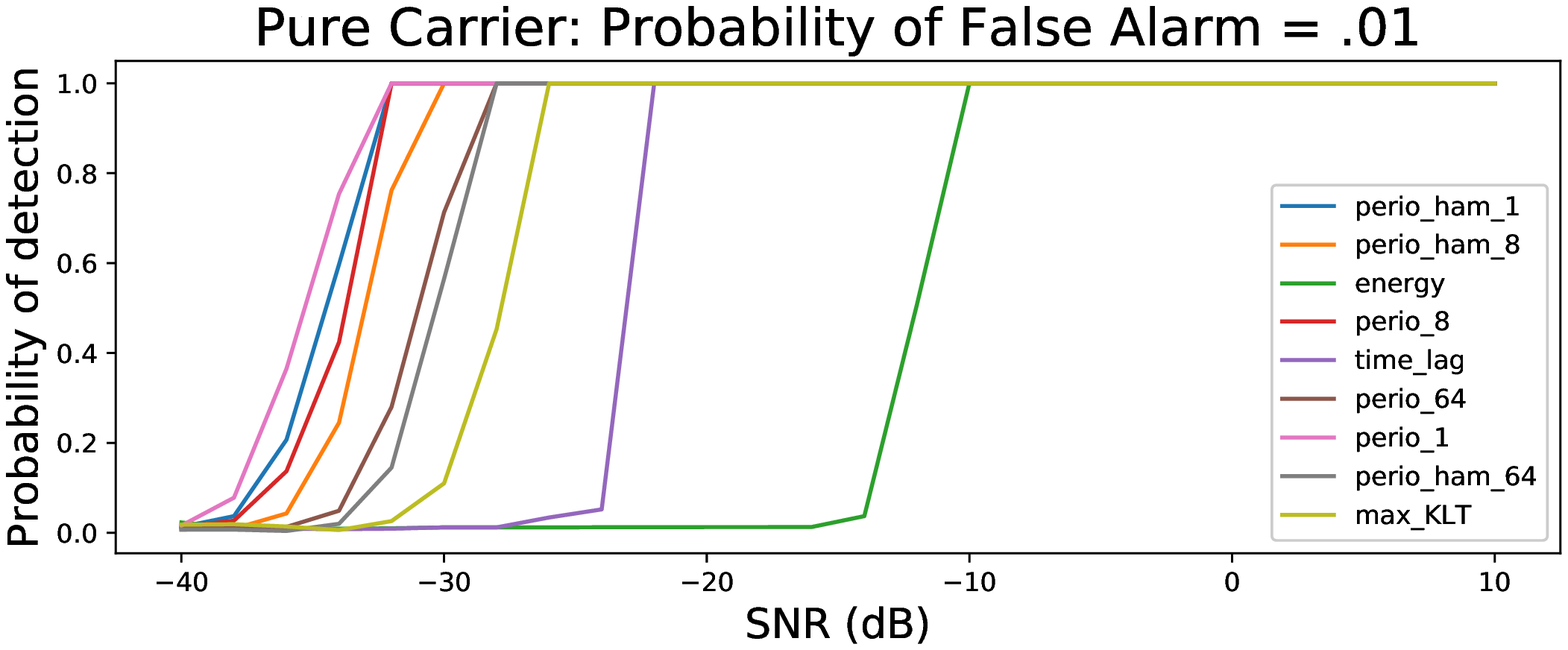}}\\

%\subfloat[]{\includegraphics[clip,width=.8\columnwidth]{./figures/chirp}}\\
\subfloat[]{\includegraphics[clip,width=.8\columnwidth]{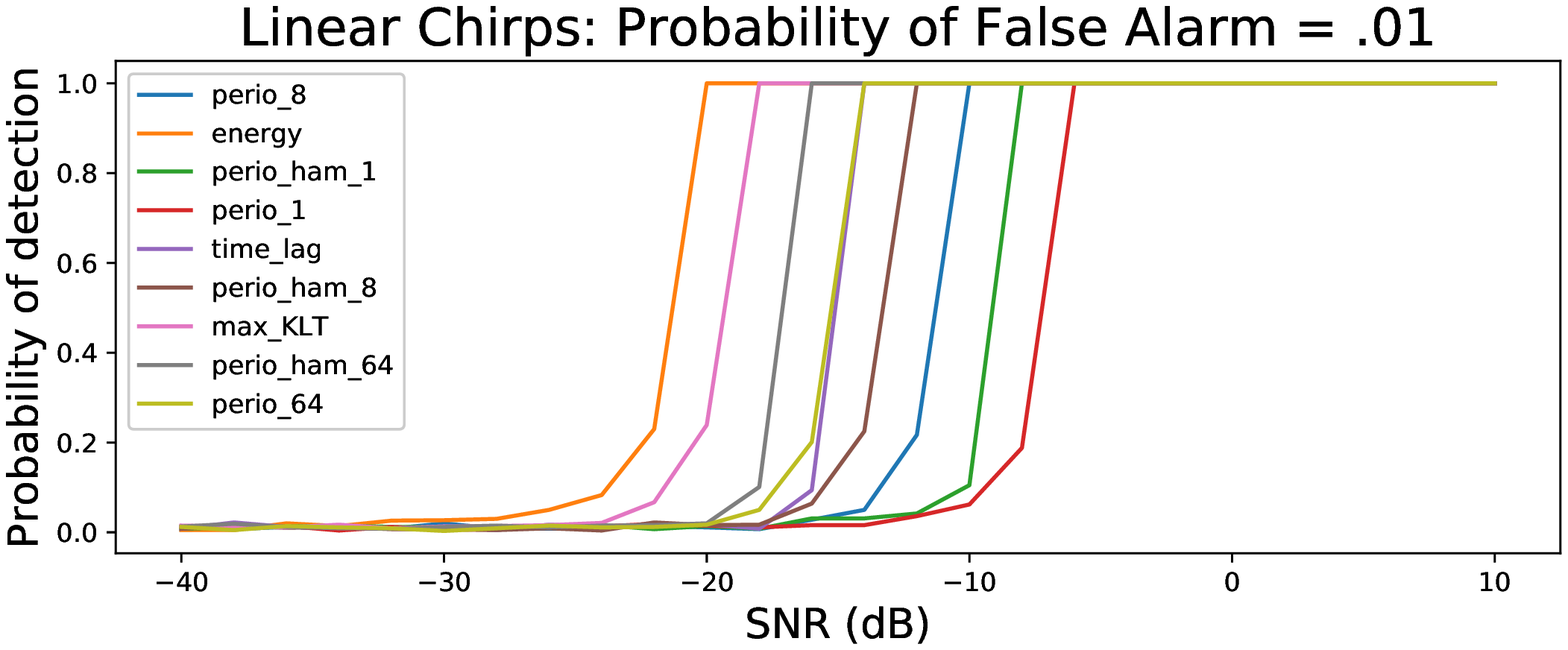}}\\

%\subfloat[]{\includegraphics[clip,width=.8\columnwidth]{./figures/BPSK}}\\
\subfloat[]{\includegraphics[clip,width=.8\columnwidth]{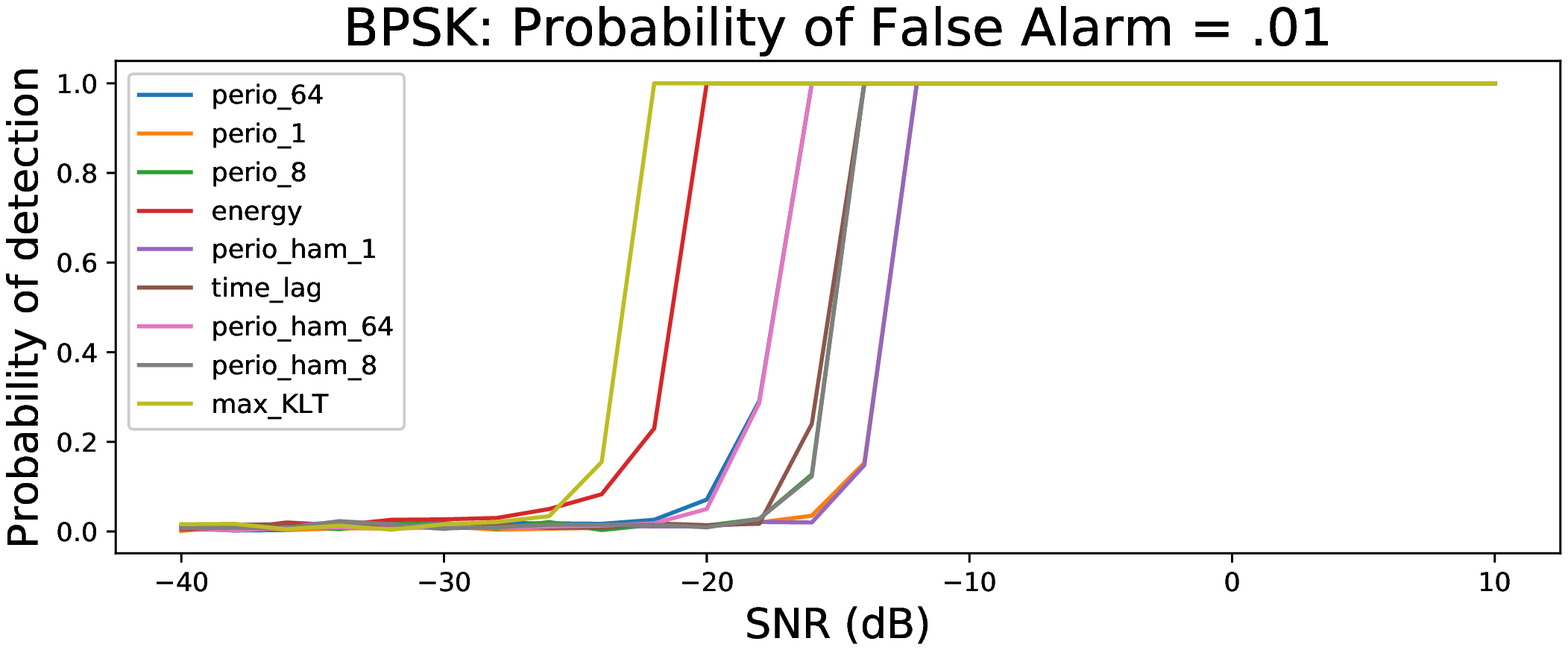}}\\

%\subfloat[]{\includegraphics[clip,width=.8\columnwidth]{./figures/bpsk+ham}}
\subfloat[]{\includegraphics[clip,width=.8\columnwidth]{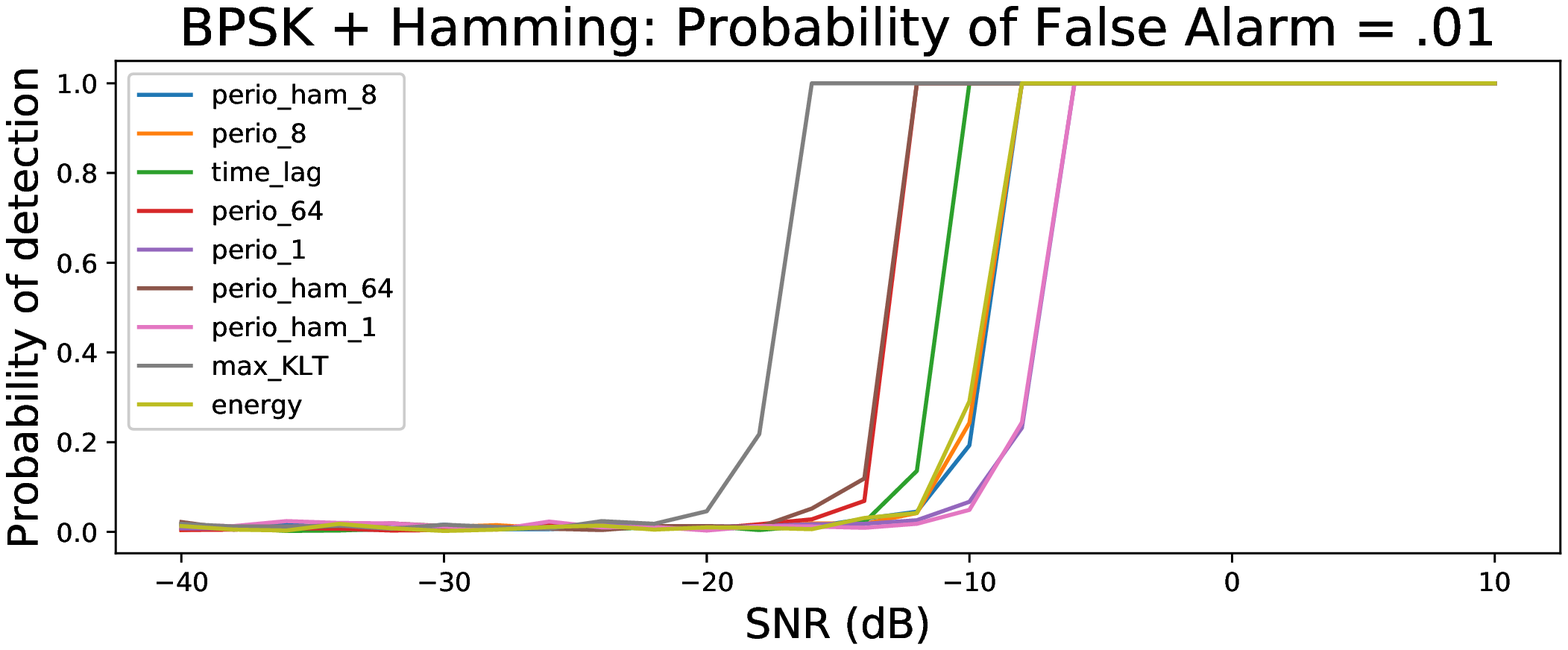}}
\caption{Probability of Detection as a function of the SNR for the considered detectors, for a given probability of False Alarm $P_{fa} = 0.01$. (a) Pure carrier model. (b) Linear chirp model. (c) BPSK - baud rate = 1/8 sampling frequency. (d) Windowed BPSK - baud rate = 1/8 sampling frequency.}
\label{fig:simu_results}
\end{figure}

Figure \ref{fig:simu_results} reports the results of this analysis. Each plot represents the Probability of Detection $P_d$ associated with each detector (figures \ref{fig:simu_results}.(a)-(d)) for a given probability of False Alarm $P_{fa} = 0.01$ for 4 signal types (pure carrier, linear chirp, BPSK and windowed BPSK). The data models parameters\footnote{except the amplitude term $A$ defining the SNR of the model for $\sigma^2=1$} are randomly generated following: $f_0\sim\mathcal{U}_{[0,1]}$, $d\sim\mathcal{U}_{[-\frac{2}{N},\frac{2}{N}]}$, $\phi\sim\mathcal{U}_{[0,1]}$, where $\mathcal{U}_{[a,b]}$ is the uniform distribution over the range $[a,b]$.

The algorithms parameterizations are as follow:
\begin{itemize}[noitemsep,nolistsep]
\item \emph{time\_lag}:\\
Time-lagged-based detection evaluated over $N-1$ time-lags.
\item \emph{perio\_\{1,8,64\}}:\\
Fourier-based detection evaluated over $N/\{1,8,64\}$ frequency bins and averaged over $\{1,8,64\}$ non-overlapping windows.
\item \emph{perio\_ham\_\{1,8,64\}}:\\
Same as above, applying an Hamming window prior to the Fourier Transform.
\item \emph{max\_KLT}:\\
Auto-correlation matrix of dimension $M=64$, averaged of $N-M$ overlapping windows.
\end{itemize}

\subsection{Simulation results}

\begin{itemize}[noitemsep,nolistsep]
\item All detectors behave efficiently for SNR $>$ -10 dB.

\item The \emph{Fourier-based detectors} outperform all detectors when applied to pure carrier signals, as they are assimilated to near-optimum matched filters in this context.

\item Linear chirps do not respond similarly with the \emph{Fourier-based detectors} as their spectra undergo frequency smearing after data integration. Shorter integration durations ($N/8$ or $N/64$) present lower spectral smearing and higher detection performances.

\item The \emph{Time-lag-based detector} presents average detection performances on all data models, failing to detect auto-correlation at low SNR.

\item The \emph{KLT detector} dominates the other detectors on wide-band transmissions.
\end{itemize}

\section{Real Data processing}
\label{sec:realdata}

To validate the theoretical results shown in the previous section, the detectors are tested on real data collected with the Green Bank telescope in West Virginia (USA). This experiment compares the recovery of Radio Frequency Interference (RFI) on $\approx$ 3 MHz-wide channelized baseband data. A software fine channelization splits the the data into 3 / 128 MHz $\approx$ 23 kHz channels.

\begin{figure}
\begin{centering}
%\centering
%\subfloat[]{\includegraphics[width=.7\columnwidth, clip]{./figures/14.eps}}\\
\subfloat[]{\includegraphics[width=.7\columnwidth, clip]{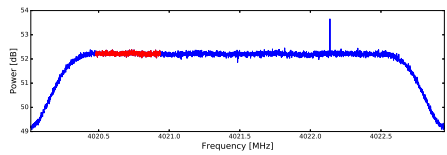}}\\
%\subfloat[]{\includegraphics[clip,width=.7\columnwidth]{./figures/Energy}}\\
\subfloat[]{\includegraphics[clip,width=.7\columnwidth]{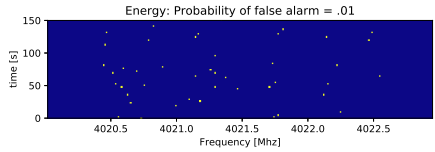}}\\
%\subfloat[]{\includegraphics[clip,width=.7\columnwidth]{./figures/KLT}}\\
\subfloat[]{\includegraphics[clip,width=.7\columnwidth]{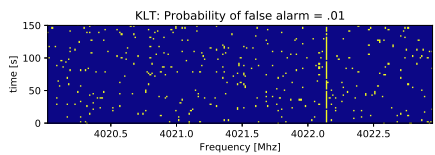}}\\
%\subfloat[]{\includegraphics[clip,width=.7\columnwidth]{./figures/perio_8}}\\
\subfloat[]{\includegraphics[clip,width=.7\columnwidth]{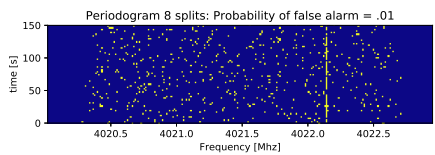}}\\
%\subfloat[]{\includegraphics[clip,width=.7\columnwidth]{./figures/perio_ham_8}}
\caption{GBT data featuring a narrow band signal. (a) Power Spectral Density evaluated over 5 min. Following binary spectrograms resulting from the \emph{Energy detector} (b), the \emph{KLT detector} (c), and the \emph{Fourier-based detector} with 8 non-overlapping spectra average (d).
%and the \emph{windowed Fourier-based detector} with 8 non-overlapping spectra average (e).
}
\label{fig:realdata1}
\end{centering}
\end{figure}

\begin{figure}
\begin{centering}
%\centering
%\subfloat[]{\includegraphics[width=.7\columnwidth,clip]{./figures/35.eps}}\\
\subfloat[]{\includegraphics[width=.7\columnwidth,clip]{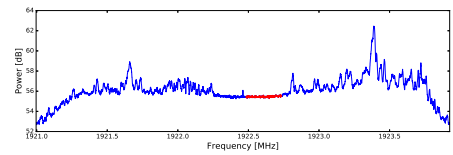}}\\
%\subfloat[]{\includegraphics[clip,width=.7\columnwidth]{./figures/b_Energy}}\\
\subfloat[]{\includegraphics[clip,width=.7\columnwidth]{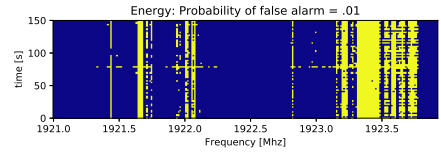}}\\
%\subfloat[]{\includegraphics[clip,width=.7\columnwidth]{./figures/b_KLT}}\\
\subfloat[]{\includegraphics[clip,width=.7\columnwidth]{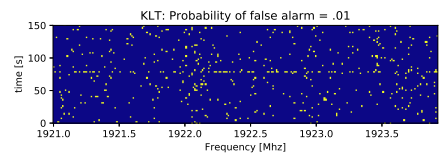}}\\
%\subfloat[]{\includegraphics[clip,width=.7\columnwidth]{./figures/b_perio_8}}\\
\subfloat[]{\includegraphics[clip,width=.7\columnwidth]{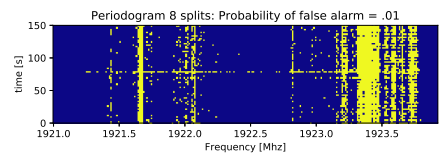}}\\
%\subfloat[]{\includegraphics[clip,width=.7\columnwidth]{./figures/b_perio_ham_8}}
\caption{GBT data featuring wide band signals. (a) Power Spectral Density evaluated over 5 min. Following binary spectrograms resulting from the \emph{Energy detector} (b), the \emph{KLT detector} (c), and the \emph{Fourier-based detector} with 8 non-overlapping spectra average (d).
%and the \emph{windowed Fourier-based detector} with 8 non-overlapping spectra average (e).
}
\label{fig:realdata2}
\end{centering}
\end{figure}

Figures \ref{fig:realdata1}.(a) and \ref{fig:realdata2}.(a) show the estimated Power Spectral Density of the two data sets integrated over 2.5 minutes. In red has been highlighted the ``eyeballed'' empty parts of the spectra used to estimate a detection threshold corresponding to a Probability of False Alarm $P_{fa}$ = 0.01. The ``roll-off'' sections on both ends of both spectra correspond to the bandpass of the 3-MHz filter bank.

The data set \#1 on figure \ref{fig:realdata1} is composed of a single narrow band transmission. The data set \#2 on figure \ref{fig:realdata2} is more complex and composed of various wide-band signals.

The following binarized spectrograms \{\ref{fig:realdata1},\ref{fig:realdata2}\}.(b-c-d) are the results detection results of the \emph{energy detector}, the \emph{KLT detector}, 	and the \emph{Fourier-based detector} with 8 non-overlapping integrations. Blue pixels correspond to a non-detection, yellow pixels correspond to a detection.

\begin{itemize}[noitemsep,nolistsep]
\item Comments on Figure \ref{fig:realdata1}\\
The \emph{Energy detector} globally fails at recovering any feature due to the low SNR of the signal.
The \emph{KLT detector} properly retrieves the narrow band signal and additional noise outliers, and is not affected by the band-pass response of the data thanks to its power normalization.
The two variations of the \emph{Fourier-based detector} also recover the signal and noise outliers over the whole data span.
%No detection is identified on the edges of the frequency spectra due to the attenuation from the filter bank band-pass response.

\item Comments on Figure \ref{fig:realdata2}\\
All detectors detect a broadband feature at $t \approx$ 80s), possibly an instrumental artifact.
The \emph{Energy detector} recovers the various wide band features with only a few noise outliers. The \emph{KLT detector} fails at recovering any signal due to weak auto-correlated features. Both variations of the \emph{Fourier-based detector} show performances comparable to the \emph{Energy detector} with higher low-SNR noise outlier detections.
\end{itemize}

\section{Conclusion}
\label{sec:conclusion}

Radio SETI aims at identifying intelligent and communicative civilizations in the Universe through the detection of their artificial transmissions. High sensitivity instrumentation and detection schemes are necessary in the absence of prior information concerning these transmissions. This paper compares four detection schemes for a given probability of false alarm based on the signal energy, frequency distribution and auto-correlation. The comparison is theoretically conducted on three data models before running them on real telescope data.

The analysis conducted shows a high dependence of a detector performance to the data model and the computational complexity. Detection performance can be improved through careful algorithm parametrization, and require further simulations and read data processing.

\end{document}